\documentclass[a4paper,12pt]{article}
\usepackage{amssymb,amsmath}
\usepackage{bm}%
\usepackage{accents}
\usepackage{cite}

\def\beq{\begin{equation}}
\def\eeq{\end{equation}}
\def\bea{\begin{eqnarray*}}
\def\eea{\end{eqnarray*}}
\def\nn{\nonumber}
\def\cga{\mathfrak{g}_{\ell}}

\def\ket#1{\left| #1\right\rangle}
\def\bracket#1#2{\left\langle #1 | #2 \right\rangle}
\def\NN{ {\mathbb Z}^{+} }
\def\N0{ {\mathbb Z}_{+} }
\def\v0{ |d,r) }

\def\Z{{\mathbb Z}}
\def\half{\frac{1}{2}}
\def\VM{V^{\delta,\mu}}
\def\hwv{\ket{\delta,\mu}}
\def\veck{ \underaccent{\tilde}{k} }
\def\vece{ \underaccent{\tilde}{\epsilon} }
\def\vec0{ \underaccent{\tilde}{0} }
\def\veckr#1{ \underaccent{\tilde}{k}^{(#1)} }
\def\a#1{ a(#1) }
\def\rket#1{ | #1 ) }
\def\proof#1{{\bf Proof:} #1 $\blacksquare$ \medskip}
\newtheorem{lemma}{Lemma}%[section]

\newtheorem{thm}[lemma]{Theorem}
\begin{document}

\begin{center}
{\large\bf Highest weight representations and Kac determinants for a class of conformal
Galilei algebras with central extension}\\
~~\\

{\large Naruhiko Aizawa$^1$, Phillip S. Isaac$^2$ and Yuta Kimura$^1$}\\
~~\\

$^1$ Department of Mathematics and Information Sciences, Osaka Prefecture University,
Nakamozu Campus, Sakai, Osaka 599-8531, Japan.\\
$^2$ School of Mathematics and Physics, The University of Queensland, St Lucia QLD 4072, Australia.
\end{center}

\begin{abstract}
We investigate the representations of a class of conformal Galilei algebras in one spatial
dimension with central extension. 
This is done by explicitly constructing all singular vectors within the Verma modules,
proving their completeness and then deducing irreducibility
of the associated highest weight quotient modules. A resulting classification of infinite
dimensional irreducible modules is presented. 
It is also shown that a formula for the Kac determinant is deduced from our construction of singular vectors. Thus we prove 
a conjecture of Dobrev, Doebner and Mrugalla for the case of the
Schr\"odinger algebra.
\end{abstract}

%%%%%%%%%%%%%%%%%%%%%%%%%%%%%%%%%%%%%%%%%%%%%%%%%%%%%%%%%%%%%%
%
%  Definitions
%
%%%%%%%%%%%%%%%%%%%%%%%%%%%%%%%%%%%%%%%%%%%%%%%%%%%%%%%%%%%%%%

\section{Introduction}

Conformal extensions of Galilei groups and their Lie algebras
\cite{HaPle,NdelOR,GalMas2011} are objects 
of physical and mathematical interest. In contrast to the relativistic conformal algebras there are 
some varieties of such algebras in a nonrelativistic setting even for a fixed dimension of spacetime. 
In this paper we investigate a class of conformal Galilei algebras specified by  ``spin'' $ \ell. $ 
The member corresponding to the smallest value $ \ell = 1/2 $ is the Schr\"odinger algebra \cite{Nie,Hagen} 
whose relevance has been observed in a wide range of physical systems 
\cite{
 Nie,Hagen,BurPer,Nie2,Nie3,BoShWi,HuJa,Henkel,Henkel3,Henkel2,HeUn,HaHo,ORS,
 Gho,ZhangHor,StoHen,JackPi,DuHoPalla,HaHo2,DGH,ADD,ADDS,MeStWi,NisSon,Son,BaMcG,
 DuHaHor,Gala,Gala2
}. 
Other members of ``spin-$\ell$'' class algebra with $ \ell > \half $ also play important roles 
in various fields in physics. For instance, one may encounter the algebra in classical mechanics with 
higher order time derivatives \cite{StZak,LSZ,LSZ2}, electrodynamics \cite{NdORM2}, 
nonrelativistic analogue of AdS/CFT correspondence \cite{MT,BG,ADV,BaMa,BaKu}, 
nonrelativistic spacetime and gravity \cite{DuHo,DH,HoKuNi}, 
quantum mechanical particle systems \cite{AGKP,GomKam2012}, twistors \cite{FeKoLuMas} and so on. 
Furthermore, one may find applications of the algebraic structure to mathematical
studies of topics such as systems of partial differential equation \cite{FuChe,FuChe2,ChHe} and matrix orthogonal polynomials \cite{ViZhe}. 

Despite the aforementioned fruitful physical applications, the representation theory of the 
``spin-$\ell$'' conformal Galilei algebras has not been thoroughly developed. 
Presumably this is due to the non-semisimple nature of the algebra, namely, the algebra defined in 
$d$-dimensional space is a semi-direct sum of $ sl(2) \oplus so(d) $ (maximal semisimple subalgebra) and 
an abelian ideal depending on $\ell. $ 
We remark that the abelian ideals for certain values of $(d,\ell)$ have central extensions \cite{Henkel,StZak,LSZ,MT}. 
One of the most fundamental problems in representation theory is a classification of 
irreducible representations. 
For the ``spin-$\ell$'' class of the conformal Galilei algebra this problem has been
solved only in certain cases. 
Irreducible representations of lowest weight type for $ \ell = 1/2 $ and $ d = 1, 2, 3 $ are 
classified in \cite{DoDoMr,Mrugalla}. 
Two of the present authors gave the list of all possible irreducible representations of highest weight 
type for $ \ell = 1 $ and $ d = 2 $ \cite{AiIs}.  
Although the classification problem of irreducible representations for other pairs of $ (d, \ell) $ is still open, 
these works show that the standard techniques for semisimple Lie algebras such as
triangular decomposition, Verma modules and 
singular vectors can be applied to this particular class of non-semisimple Lie algebra. 
Motivated by this fact, in the current work we shall undertake further investigation of the classification problem.  
A goal of the present work is to give a list of all possible highest weight irreducible modules of  
the algebra with a central extension for $ d = 1 $ and arbitrary half-integer $\ell. $ 
This will be done by constructing singular vectors explicitly in Verma modules. 

After the classification of irreducible modules, we derive the formula of the Kac determinant. 
Usually the Kac determinant is calculated in order to detect the existence of singular vectors. 
Here we are able to give the Kac determinant as a corollary of our explicit construction of singular vectors. 
In other words, we would like to show that our construction of singular vectors leads to
an explicit formula for the Kac determinant. We include this because the calculation of the Kac determinant
for the ``spin-$\ell$'' conformal 
Galilei algebra based on its definition is rather difficult. Even for the simplest example
of $ \ell = 1/2,\; d = 1, $ 
only a conjectured form of the Kac determinant has been given\cite{DoDoMr}. 
We give the Kac determinant for the $d = 1$ algebra with arbitrary half-integer $\ell.$ 
As a result we show that the conjectured formula presented in \cite{DoDoMr} is true. 

We organize the paper as follows. In the next section we give the definition of $ d = 1 $ algebra and 
fix our conventions. We then introduce the Verma modules and give the singular vectors in \S \ref{Sec:SV}. 
It is shown that a Verma module has precisely one singular vector if it exists. Then follows a classification 
of irreducible modules of highest weight type. In \S \ref{Sec:KD} we derive the formula of
the Kac determinant. 
%\S \ref{Sec:CR} is concluding remarks. 

\section{Structure of the conformal Galilei algebras, conventions and preliminaries}

For\footnote{Note that throughout the paper we use the following notation for the sets of 
non-negative and positive integers respectively: $ \N0 = \{ \ 0, 1, 2, \cdots \ \},\ \  
\NN = \{ \ 1, 2, \cdots \ \}.  $ } $ \ell\in \half \NN$, the ``spin-$\ell$'' class
\cite{NdelOR} of the conformal Galilei algebra
(without central extension) defined in (1+1)-dimensional spacetime has a basis given by 
$$ 
\left\{ C, D, H, P_n \ |\ n = 0, 1, 2, \cdots, 2 \ell\right\}. 
$$
Our purpose is to study the representation theory of the centrally extended algebra, which
we denote $\cga$.
It is well known \cite{MT} that in the case where $\ell$ takes on integer
values, no such central extension exists in the case of one spatial dimension. It is also
known \cite{DoDoMr,Mrugalla} that the irreducible highest weight representations in the case $\ell = 1/2$
algebra without central extension reduce to those of the $sl(2)$ subalgebra spanned by
$C$, $D$ and $H$. Therefore, in this article we focus only on the algebra with central
extension, i.e., $\ell = 1/2, 3/2, 5/2, 7/2,\ldots$, i.e. the odd positive half-integers.

To this end, the non-zero defining relations of the algebra $\cga$ under consideration are
given by 
\[
  \begin{array}{lll}
     [D, H] = 2H, & [D, C] = -2C, & [C, H] = D, \\[5pt]
     [H, P_n] = -n P_{n-1},& [D, P_n] = 2(\ell-n) P_n, & [C, P_n]= (2\ell-n) P_{n+1}, \\[5pt]
     [P_m, P_n] = I_{m,n}  M. & & 
  \end{array}
\]
Here $M$ is the central extension (so $[M,a]=0$ $\forall a\in \cga$), and the structure
constants $I_{m,n}$ form an antisymmetric tensor that can be determined by consistency
with the Jacobi identity. In fact, from the Jacobi identity we first
observe that the $I_{m,n}$ can only be nontrivial if $m+n=2\ell$, in which case the following linear
equations must be satisfied:
$$
(2\ell-n)I_{2\ell-n-1,n+1} = -(n+1)I_{2\ell-n,n},\ \ n=0,1,2,\ldots,2\ell-1.
$$
Solving these equations gives
$$
I_{m,n} = \delta_{m+n, 2\ell}\, (-1)^{m+\ell+\frac12}m!n!\beta_\ell,
$$
where $\beta_\ell$ is an arbitrary constant that depends on $\ell$. For convenience we set
\beq
I_m = (-1)^{m+\ell+\frac12}m!(2\ell-m)!, \label{strcon-I}
\eeq
and herein use the relation
\beq
[P_m,P_n] = \delta_{m+n, 2\ell}\, I_m M, \label{centrel}
\eeq
ignoring the factor $\beta_\ell$ by absorbing it into the central extension $M$. We note
that this agrees, up to an overall factor, with the structure constants used in
\cite{GalMas2011} and \cite{GomKam2012}.

One can immediately deduce a triangular decomposition of $ \cga $:
\bea
 & &  \cga^+ = \{ \ H, \ P_0, \ P_1, \ \cdots, \ P_{\ell-\half} \ \} \\
 & &  \cga^0 = \{ \ D, \ M \ \} \\
 & &  \cga^- = \{ \ C, \ P_{\ell+\half}, \ P_{\ell+\frac{3}{2}}, \ \cdots, \ P_{2\ell}. \ \}
\eea
We remark that $ \cga^{\pm} $ are Abelian for $ \ell = \half $ and non-Abelian for $ \ell \geq \frac{3}{2}. $ 

\section{Singular vectors and irreducible modules}
\label{Sec:SV}

Our approach for finding all irreducible highest weight modules of $\cga$ is according to
the following procedure, which generalises that of \cite{DoDoMr} for the case
$\ell=1/2$ (Schr\"odinger algebra).

\begin{enumerate}
\item Define a highest weight vector of the Verma module in a basis on which $\cga^0$ is
diagonal, and give the action of the generators in $\cga$, and hence a basis
for the Verma module.
\item Determine a complete set of singular vectors in the Verma module.
\item Quotient out all highest weight submodules generated by the singular vectors. The
remaining quotient module is then irreducible.
\end{enumerate}
%%%%%%%%%%%%%%%%%%%%%%%%%%%%%%%%%%%%%%%%%%%%%%%%%%%%%%%%%%%%%%%%
%
%  Verma module
%
%%%%%%%%%%%%%%%%%%%%%%%%%%%%%%%%%%%%%%%%%%%%%%%%%%%%%%%%%%%%%%%%
%

\subsection{Verma module and basis}

For fixed, real values of $\delta$ and $\mu$, let $\ket{\delta,\mu}$ be a highest weight vector in the Verma module, such that
\[
  D \hwv = \delta \hwv, \qquad M \hwv = \mu \hwv, \qquad 
  X \hwv = 0, \quad X \in \cga^+.
\]
The Verma module itself, denoted $\VM$, is determined by 
$ U(\cga^-) \hwv, $ with $U(\cga^-)$ being the universal enveloping algebra of
$\cga^-.$ We then have a natural basis of $\VM$ given by
\[
 \left\{ \left. C^h \prod_{j=0}^{\ell-\half} P_{\ell+\half + j}^{k_j} \hwv\ \right| \quad 
  h, k_0,k_1,\ldots,k_{\ell-\frac12} \in \N0 \right\}.
\]
It is easily seen that each basis vector is simultaneously an eigenvector of $ M $ and
$D.$ The eigenvalue of $M$ is always $\mu$, since $M$ is central. Using the commutation
relations of $D$ with powers of $C$ and $P_j$, namely
\beq
[D,C^h]=-2hC^h,\quad [D,P_n^k] = 2k(n-\ell)P_n^k, \label{Drel}
\eeq
the eigenvalue of $D$ corresponding to basis vector $\displaystyle{C^h \prod_{j=0}^{\ell-\half}
P_{\ell+\half + j}^{k_j} \hwv}$ is found to be
$$ 
\delta - 2h - \sum_{j=0}^{\ell-\half} (2j+1) k_j.
$$ 
This suggests that we introduce the notion of {\em level} within $\VM$. 
We define the level $m$ by 
\beq
  m = 2h + \sum_{j=0}^{\ell-\half} (2j+1) k_j. \label{Level-def}
\eeq
For a fixed value of $m$ we take $ h, k_1, k_2, \cdots, k_{\ell-\half} $ as independent
variables, since
\beq
k_0 = m - 2h - \sum_{j=1}^{\ell-\half} (2j+1) k_j, \label{k-not}
\eeq
and in particular,
\beq
m - 2h - \sum_{j=1}^{\ell-\half} (2j+1) k_j\geq 0. \label{label-suff}
\eeq

We then find it convenient throughout the paper to denote the basis vectors at level $m$
by $\ket{h, \veck ; m}$, where
\beq
  \ket{h, \veck ; m} =  C^h P_{\ell+\half}^{m - 2h - \sum_{j=1}^{\ell-\half} (2j+1)
k_j}\prod_{j=1}^{\ell-\half} P_{\ell+\half + j}^{k_j} \hwv,
\label{mbasis}
\eeq
with $ \veck = (k_1, k_2, \cdots, k_{\ell-\half}). $ Note that the above expression
appears cumbersome, so
to facilitate readability of the equations throughout the paper, where possible we write $k_0$ (which is always the power of the
generator $P_{\ell+\frac12}$ occurring in the basis vectors) instead of the full expression in terms of
$m$, $h$ and $k_1,\ldots,k_{\ell-\frac12}$ as determined by equation (\ref{k-not}). We
then have expressions such as
$$
\ket{h, \veck ; m} =  C^h\prod_{j=0}^{\ell-\half} P_{\ell+\half + j}^{k_j} \hwv
$$
making perfect sense by keeping equation (\ref{k-not}) in mind. 

The Verma module $ \VM $ is a graded module with respect to the vector space decomposition
over $m$,
\[
  \VM = \bigoplus_{m \in \N0} \VM_m, 
\]
where $ \VM_m $ is the space spanned by the vectors $ \ket{h, \veck ; m} $ for fixed $m.$ 

%%%%%%%%%%%%%%%%%%%%%%%%%%%%%%%%%%%%%%%%%%%%%%%%%%%%%%%%%%%%%%%%
%
%  Formula of singular vector
%
%%%%%%%%%%%%%%%%%%%%%%%%%%%%%%%%%%%%%%%%%%%%%%%%%%%%%%%%%%%%%%%%
%
\subsection{Singular vectors}

As mentioned in the outline at the start of this section, we proceed by stating an
explicit form of singular vectors, and then show that these constitute a complete 
set of singular vectors in $\VM$. We remind the reader that the vector $\hwv$ at level 0
is not considered to be a singular vector.

\begin{thm} \label{thm1}
If $ 2 \delta - 2(q-1) + (\ell+\half)^2 = 0 $ for $ q \in \NN $ then the following 
is a singular vector at level $ 2q: $
\beq
   \ket{u_{2q}} = ( \alpha_{\ell} \mu \, C - P_{\ell+\half}^2 )^q \hwv \in \VM_{2q}
   \label{SV}
\eeq
where $ \alpha_{\ell} = 2 ( (\ell-\half)! )^2. $  
\end{thm}
\proof{
Using the relations 
\bea
& & [P_j, \, \alpha_{\ell} \mu \, C - P_{\ell+\half}^2] 
= - (2\ell -j) \alpha_{\ell} \mu P_{j+1}, 
\quad
0 \leq j \leq \ell - \frac{3}{2},
\\[3pt]
%%% & & [P_{\ell-\half}, \, \alpha_{\ell} \mu \, C - P_{\ell+\half}^2] 
%%% = - \left( \ell+\half \right) \alpha_{\ell} \mu P_{\ell+\half} 
%%% + 2 \left( \ell+\half \right) ! \left( \ell - \half \right) ! M P_{\ell+\half},
%%%
%%% Mar 30 NA
%%%
& & [P_{\ell-\half}, \, \alpha_{\ell} \mu \, C - P_{\ell+\half}^2] 
= \left( \ell+\half \right) \alpha_{\ell} (M-\mu) P_{\ell+\half}, 
\\[3pt]
& & [H, \, \alpha_{\ell} \mu \, C - P_{\ell+\half}^2] 
= -\alpha_{\ell} \mu D + 2 \left( \ell + \half \right) P_{\ell+\half} P_{\ell-\half} 
- \left( \left( \ell + \half \right) ! \right)^2 M,
\eea
one may use a simple inductive argument on $ q $ to prove the following:
\bea
& & P_j \ket{u_{2q}} = 0, \quad 0 \leq j \leq \ell - \half,
\\
& & H \ket{u_{2q}} = 
- \frac{q}{2} \alpha_{\ell} \mu 
\left\{ \ 2 \delta - 2(q-1) + \left( \ell + \half \right)^2 \ \right\} 
\ket{ u_{2(q-1)} }.
\eea
It is straightforward to verify 
\[
D \ket{u_{2q}} = (\delta - 2 q ) \ket{ u_{2q} }.
\]
The result of the theorem follows.
}
%%%%%%%%%%%%%%%%%%%%%%%%%%%%%%%%%%%%%%%%%%%%%%%%%%%%%%%%%%%%%%%%
%
%  Uniqueness
%
%%%%%%%%%%%%%%%%%%%%%%%%%%%%%%%%%%%%%%%%%%%%%%%%%%%%%%%%%%%%%%%%
%

Each singular vector given by Theorem \ref{thm1} is a homogeneous vector in the Verma
module. That is, for a given (even) level $m$, the singular vector $ \ket{u_m} $ can be
expressed as a linear combination of the basis vectors of $ \VM_m: $ 
\beq
\ket{u_m} = \sum_{h,\veck} \a{h,\veck} \ket{h, \veck ; m}, \label{svsum}
\eeq
where the sum on $ h, \veck $ is over all possible values satisfying (\ref{label-suff}). 

We aim to show that the conditions $ X \ket{u_m} = 0 $ for any $ X \in \cga^+ $ 
determine all the coefficients $ \a{h,\veck} $ uniquely up to overall constant. From this
we are then able to deduce that the singular vectors given by Theorem \ref{thm1} are the
only singular vectors, and hence constitute a complete set.

We start with some useful preliminary results involving the defining relations, and remind
the reader about the relation (\ref{centrel}), and the definition of the structure
constant $I_m$ given in equation (\ref{strcon-I}).
\begin{lemma} \label{lemma2}
The following commutation relations hold on $U(\cga)$:
\begin{eqnarray}
  & & 
   [P_j, C^h] = \sum_{n=1}^{ \min\{ h,2\ell-j \} }(-1)^n n! 
    \begin{pmatrix}
      h \\ n
    \end{pmatrix} 
    \begin{pmatrix}
     2\ell-j \\ n
    \end{pmatrix} 
    C^{h -n} P_{j+n}, \label{rel-PC}
  \\[3pt]
  & & 
   [P_j, P_{2\ell-j}^k] = k I_j M P_{2\ell-j}^{k-1}, 
   \label{rel-PP}
  \\[3pt]
  & & 
   [H, C^h] = -h C^{h-1} D + h(h-1) C^{h-1},  \label{rel-HC}
  \\[3pt]
  & &
   [H, P_n^k] = -kn P_n^{k-1} P_{n-1} 
    + \delta_{n,\ell+\half} \, \half k (k-1) \left( \left( \ell + \half \right) ! \right)^2 M P_n^{k-2}. 
    \label{rel-HP}
\end{eqnarray}
\end{lemma}
\proof{
Each relation can be proved using a straightforward inductive argument. The details are
unenlightening and omitted.
}

Before proceeding, we introduce some useful notation to assist in keeping track of shifts in the state
labels. Let $ \vece_j $ be the row vector in $ {\mathbb R}^{\ell-\half} $ with 1 in the
$j$-th entry and 0 elsewhere, i.e.,  
\[
  \vece_j = (0, \cdots, 0,1,0 \cdots, 0).
\]
We also use the notation
\[
  \veck^{(r)} = (k_1, k_2, \cdots, k_r, 0, \cdots, 0) \in {\mathbb R}^{\ell-\half}.
\]
We adopt the convention that $ \vece_0 = (0, \cdots, 0) = \veckr{0}. $ 

\begin{lemma} \label{lemma3} 
The action of $ \cga^+\cup \cga^0$ on $ \ket{h,\veck; m}$ is given by the following expressions: 
$$
M \ket{h,\veck; m} = \mu\ket{h,\veck; m},
$$
\beq
 D \ket{h,\veck; m} = (\delta - m) \ket{h, \veck; m}, 
   \label{Donket}
\eeq
\begin{eqnarray}
 & & P_{\ell-\half-a} \ket{h,\veck;m} 
 \nn \\ 
 & & \quad = 
  \sum_{0 \leq n \leq a} (-1)^n k_{a-n}I_{\ell-\half-a+n} 
   n ! \begin{pmatrix} h \\ n \end{pmatrix}
       \begin{pmatrix} \ell+\half+a \\ n \end{pmatrix} \mu 
       \ket{h-n, \veck-\vece_{a-n}; m-2a-1}
 \nn \\[3pt]
 & & \quad + \sum_{a+1 \leq n} (-1)^n n! 
       \begin{pmatrix} h \\ n \end{pmatrix}
       \begin{pmatrix} \ell+\half+a \\ n \end{pmatrix} 
       \ket{h-n, \veck+\vece_{n-a-1}; m-2a-1},
 \label{Ponket}
\end{eqnarray}
where $ a = 0, 1, \cdots, \ell-\half. $ 
\begin{eqnarray}
 H \ket{h,\veck;m} &=& h (m-h-\delta-1) \ket{h-1,\veck;m-2}
 \nn \\[3pt] 
   &+& \half k_0 (k_0-1) \left( \left( \ell+\half \right)! \right)^2 \mu \ket{h,\veck;m-2}
 \nn \\[3pt] 
   &-& \sum_{j=1}^{\ell-\half} k_j \left( \ell+\half+j \right) \ket{h, \veck+\vece_{j-1}-\vece_j;m-2}.
   \label{Honket}
\end{eqnarray}
\end{lemma}
\proof{
Firstly, the action of $M$ is trivial and only included for completeness. 
By application of the commutation relations (\ref{Drel}) and those of Lemma
\ref{lemma2}, the remaining actions given in this Lemma can easily be deduced, so we omit details.
}

By the result of Lemma \ref{lemma3}, in particular equations (\ref{Ponket}) and
(\ref{Honket}), one can then calculate the action of 
$\cga^+$ on $\ket{u_m}.$ We thus arrive at the following technical theorem
which is crucial to our discussion. 

\begin{thm} \label{thm4}
In order for the vector given in equation (\ref{svsum}), (reproduced here)
$$
\ket{u_m} = \sum_{h,\veck} \a{h,\veck} \ket{h, \veck ; m},
$$
to be a singular vector, $m$ must be even, in which case the
coefficients $ \a{h,\veck}$ are unique up to an overall factor.
\end{thm}
{\bf Proof:}
We determine conditions on $\a{h,\veck}$ such that $X\ket{u_m}=0,$ $\forall$ $X\in\cga^+.$

We first note that the largest possible value of $h$ for a given level $m$ is 
$ \lfloor \frac{m}{2} \rfloor. $ More precisely, the following choice of 
$ h, k_0, \veck $ gives the largest possible value of $h:$ 
\[
 \begin{array}{lcl}
   m = 2q \mbox{ (even) } &;  &\quad h = q, \quad k_0 = 0, \quad \veck = \vec0,
  \\[5pt]
  m = 2q + 1 \mbox{ (odd) } &; & \quad h = q, \quad k_0 = 1, \quad \veck = \vec0.
 \end{array}
\]
By (\ref{Ponket}) with $ a = 0 $ one has
\begin{eqnarray}
  P_{\ell-\half} \ket{u_m} &=& 
    \sum_{h,\veck} \a{h,\veck} \, 
    \{ \, 
      k_0 I_{\ell-\half} \mu \ket{h,\veck;m-1}
   \nn \\
   &+& \sum_{1 \leq n} (-1)^n n! 
       \begin{pmatrix} h \\ n \end{pmatrix}
       \begin{pmatrix} h + \half \\ n \end{pmatrix} 
       \ket{h-n,\veck+\vece_{n-1}; m-1} \, 
    \}
   \nn \\
   &=& 0. 
   \label{P-0}
\end{eqnarray}
We look at the coefficients of the vector $ \ket{q, \vec0; 2q-1} $ for even $m,$ and 
$ \ket{q,\vec0;2q} $ for odd $m.$ We see that there is no such vector for even $m$ and 
only one for odd $m$ with the coefficient $ \a{q,\vec0} I_{\ell-\half} \mu. $ 
Thus we have
\[
   \a{q, \, \vec0} = 
   \left\{
     \begin{array}{ll}
        \mbox{arbitrary,} & \mbox{for}\ m=2q, \\[7pt]
        0, & \mbox{for}\ m=2q+1.
     \end{array}
   \right.
\]
Next we look at the coefficients of $ \ket{h, \vec0; m-1} $ and 
obtain the recurrence relation:
\beq
  (m-2h) I_{\ell-\half} \mu \a{h, \, \vec0} - (h+1) \left( \ell + \half \right) \a{h+1, \, \vec0} = 0.
  \label{recrel-half}
\eeq
This recurrence relation is easily solved and we have
\[
   \a{h, \, \vec0} = 
   \left\{
     \begin{array}{ll}
        \left( - \alpha_{\ell} \mu \right)^{h-q} 
          \begin{pmatrix}
             q \\ h
          \end{pmatrix}
          \a{q,\vec0}, & \mbox{for}\ m=2q, \\[10pt]
        0, & \mbox{for}\ m=2q+1.
     \end{array}
   \right.
\]
This shows that $ \a{h, \vec0} $ is unique up to $ \a{q,\vec0},$ for $0\leq h\leq q-1.$ 

Now we look at the coefficients of the vector $ \ket{h, \veckr{r}; m-2} $ in 
the equation $ H \ket{u_m} = 0. $ This is done by using (\ref{Honket}) and 
gives the recurrence relation
\begin{eqnarray}
 & &
  (h+1) (m-h-2-\delta)\, \a{h+1,\, \veckr{r}} 
  + \half k_0 (k_0-1) \left( \left( \ell + \half \right) ! \right)^2 \mu \, \a{h,\, \veckr{r}}
 \nn \\[3pt]
 & & 
  \qquad 
  - \sum_{j=1}^{r} (k_j+1) \left( \ell + \half + j \right) \a{h,\, \veckr{r}- \vece_{j-1} + \vece_j}
 \nn \\[3pt]
 & & \qquad 
  - \left( \ell + \frac{3}{2} + r \right) \a{h,\, \veckr{r}-\vece_r + \vece_{r+1} } = 0.
 \label{recrel-H}
\end{eqnarray}
This relation  allows us to write $ \a{h,\, \veckr{r}+\vece_{r+1} } $ as 
a linear combination of the coefficients $\a{s,\, \veckr{j}},$ for all possible $s$ and for $j\leq r.$  
For instance, setting $ r = 0, $ the relation (\ref{recrel-H}) reads as follows
\begin{eqnarray}
 & &
  (h+1) (m-h - \delta -2) \a{h+1, \, \vec0} + \half (m-2h) (m-2h-1) \left( \left( \ell + \half \right) ! \right)^2 \mu \, 
  \a{h,  \, \vec0}
 \nn \\
 & & 
   - \left( \ell + \frac{3}{2} \right) \a{h,  \, \vece_1} = 0.
  \label{rec1} 
\end{eqnarray}
We see that $ \a{h, \, \vece_1 } $ is therefore determined once we know coefficients of the form $ \a{h,\vec0}: $ 
\[
  \a{h, \, \vece_1 } = 
   \left\{
     \begin{array}{ll}
        \mbox{unique up to } \a{q,\vec0}, & \mbox{for}\ m=2q, \\[7pt]
        0, & \mbox{for}\ m=2q+1. 
     \end{array}
   \right.
\]
Next we look at coefficients of the vector $ \ket{h-a, \veckr{a}; m-2a-1} $ in 
$ P_{\ell-\half-a} \ket{ u_m } = 0. $ 
This may be done by using (\ref{Ponket}). For $ 0 \leq n \leq a-1 $ we make the 
replacement
\[
  h \to h + n-a, \qquad k_{a-n} \to k_{a-n} + 1,
\]
and for $ a+1 \leq n \leq 2a+1, $ 
\[
  h \to h + n -a , \qquad k_{n-a-1} \to k_{n-a-1} -1,
\]
but no change for $ n = a. $ 
This leads to the following recurrence relation:
\begin{eqnarray}
  & & 
   \sum_{n=0}^{a-1} (-1)^n (k_{a-n}+1) I_{\ell-\half-a+n} n! 
   \begin{pmatrix}
      h + n -a \\ n
   \end{pmatrix}
   \begin{pmatrix}
     \ell + \half + a \\ n
   \end{pmatrix}
   \mu \, \a{h+n-a, \, \veckr{a}+\vece_{a-n}}
  \nn \\
  & & 
   + (-1)^a k_0 I_{\ell-\half} a! 
   \begin{pmatrix}
     h \\ a
   \end{pmatrix} 
   \begin{pmatrix}
     \ell+\half+a \\ a
   \end{pmatrix}
   \mu \, \a{h, \, \veckr{a}}
  \nn \\
  & & 
   + \sum_{n=a+1}^{2a+1} (-1)^n n! 
   \begin{pmatrix}
     h+n-a \\ n
   \end{pmatrix}
   \begin{pmatrix}
     \ell+\half+a \\ n 
   \end{pmatrix} 
   \a{h+n-a, \, \veckr{a}-\vece_{n-a-1}} = 0.
 \label{recrel-P}
\end{eqnarray}
The relation (\ref{recrel-P}) relates the following coefficients:
\bea
  & & 
   \a{h-a, \, \veckr{a}+\vece_a}, \ \a{h-a+1, \, \veckr{a}+\vece_{a-1}}, \ 
   \cdots, \ 
   \a{h-1,  \, \veckr{a}+\vece_1}, \ 
   \a{h,  \, \veckr{a}}, 
  \\[3pt]
  & & 
   \a{h+1,  \, \veckr{a}}, \ \a{h+2, \, \veckr{a} - \vece_1}, \ 
   \cdots, \ 
   \a{h+a+1,  \, \veckr{a}-\vece_a}.
\eea
Thus the relation (\ref{recrel-P}) allows us to write $ \a{h,\veckr{a}+\vece_a} $ as a
linear combination of coefficients of the form $\a{s,\veckr{a}}$ and $
\a{s',\veckr{a}-\vece_j} $ for some $s,s'$ and $j\leq a$.
For instance, setting $ a = 1 $ in (\ref{recrel-P}) we have
\begin{eqnarray}
 & & 
  (k_1 + 1) I_{\ell-\frac{3}{2}} \mu \, \a{h-1,\, \veckr{1}+\vece_1} 
  - k_0 I_{\ell-\half} h \left( \ell + \frac{3}{2} \right) \mu \, 
    \a{h, \, \veckr{1}}
 \nn \\
 & & 
   + \sum_{n=2}^3 (-1)^n n! 
     \begin{pmatrix}
       h + n -1 \\ n
     \end{pmatrix}
     \begin{pmatrix}
       \ell + \frac{3}{2} \\ n
     \end{pmatrix}
     \a{h+n-1, \, \veckr{1}-\vece_{n-2}} = 0.
  \label{reca=1}
\end{eqnarray}
Repeated use of this relation leads to the conclusion
\[
  \a{h, \, \veckr{1} } = 
   \left\{
     \begin{array}{ll}
        \mbox{unique up to } \a{q,\vec0}, & \mbox{for}\ m=2q, \\[7pt]
        0, & \mbox{for}\ m=2q+1.
     \end{array}
   \right.  
\]
We remark that by setting $ k_1 = 0, $ the relation (\ref{reca=1}) gives the
connection between  
$ \a{h,\vece_1} $ and $ \a{h,\vec0}, $ in a similar way to (\ref{rec1}):
\beq 
   I_{\ell-\frac{3}{2}} \mu \, \a{h-1, \,\vece_1} - k_0 I_{\ell-\half} h \left( \ell + \frac{3}{2} \right) \mu \, \a{h,\,\vec0}
   + 2 \begin{pmatrix}
         h+1 \\ 2
       \end{pmatrix}
       \begin{pmatrix}
         \ell + \frac{3}{2} \\ 2
       \end{pmatrix} 
   \a{h+1,\vec0} = 0.
   \label{reck1=0}
\eeq
This also gives $ \a{h,\vece_1} $ but it must be compatible with the previous computation from (\ref{rec1}). 
For odd level $m$ both (\ref{rec1}) and (\ref{reck1=0}) give $ \a{h, \vece_1} = 0 $ so that 
they are compatible. For even $ m = 2q $ with the aid of (\ref{recrel-half}) 
it follows from (\ref{rec1}) that 
\[
  \a{h, \, \vece_1} = 
  \left( \ell-\half \right) (q-h) (q-h-1) \, \alpha_{\ell} \mu \, \a{h, \, \vec0},
\]
and from (\ref{reck1=0}) 
\[
  \left( \ell + \frac{3}{2} \right) \a{h, \, \vece_1} =  (q-h) 
  \left\{ \,
    h + \delta - 2(q-1) + \left( q-h-\half \right) \left( \ell+\half \right)^2 
  \,\right\}
  \alpha_{\ell} \mu \, \a{h, \, \vec0}.
\]
Removing $ \a{h,\vece_1} $ from the two above equations, one obtains the condition for $
\delta $ found in Theorem \ref{thm1}:
\beq
  2 \delta - 2 (q-1) + \left( \ell + \half \right)^2 = 0. 
  \label{delta-cond}
\eeq

  We now return to the relation (\ref{recrel-H}) and set $ r = 1. $ 
Then we see that
\[
  \a{h, \, \veckr{1}+\vece_2 } = 
   \left\{
     \begin{array}{ll}
        \mbox{unique up to } \a{q,\vec0}, & \mbox{for}\ m=2q, \\[7pt]
        0, & \mbox{for}\ m=2q+1.
     \end{array}
   \right.
\]
By the relation (\ref{recrel-P}) with $ a=2, $ it is not difficult to see
\[
  \a{h, \, \veckr{2} } = 
   \left\{
     \begin{array}{ll}
        \mbox{unique up to } \a{q,\vec0}, & \mbox{for}\ m=2q, \\[7pt]
        0, & \mbox{for}\ m=2q+1.
     \end{array}
   \right.
\]
Repeating this it may be proved that
\[
  \a{h, \, \veck } = 
   \left\{
     \begin{array}{ll}
        \mbox{unique up to } \a{q,\vec0}, & \mbox{for}\ m=2q, \\[7pt]
        0, & \mbox{for}\ m=2q+1,
     \end{array}
   \right.
\]
which is enough to complete the proof of the theorem.
$\blacksquare$ \medskip

We have shown that $ \a{h, \, \veck} = 0 $ for odd $ m $ and uniquely determined up to 
$ \a{q,\,\vec0} $ for even $m.$ Therefore, only at even levels $m=2q$ do singular vectors
exist in $ \VM, $ with $ \delta $ satisfying (\ref{delta-cond}). Hence we may proceed with
a classification of irreducible highest weight modules of $\cga$.

%%%%%%%%%%%%%%%%%%%%%%%%%%%%%%%%%%%%%%%%%%%%%%%%%%%%%%%%%%%%%%%%
%
%  Classification of Irr modules
%
%%%%%%%%%%%%%%%%%%%%%%%%%%%%%%%%%%%%%%%%%%%%%%%%%%%%%%%%%%%%%%%%
%

\subsection{Classification of irreducible highest weight modules }

\medskip
 We have shown that $ \VM $ with $ \delta $ satisfying (\ref{delta-cond}) has 
precisely one singular vector given by (\ref{SV}). 
It follows that $ \VM $ contains the invariant submodule
\[
  I^{\delta, \mu} = U( \cga^- ) \ket{u_{2q} }. 
\]
Now we show that there are no singular vectors in the quotient module $ \VM / I^{\delta,\mu} $ 
so that the quotient module is irreducible. 
Let $ \rket{0} $ be the highest weight vector in $ \VM / I^{\delta,\mu}. $ 
A basis of $ \VM / I^{\delta,\mu} $ is then given by 
\[
 \left\{ \left. C^h \prod_{j=0}^{\ell-\half} P_{\ell+\half + j}^{k_j} \rket{0} \quad \right| \quad 
  h= 0, 1, \cdots, q-1, \ k_j \in \N0 \right\}.
\]
For a fixed level $m$ we denote a basis vector by $ \rket{h, \veck; m} $ and a 
singular vector by 
\[
   \rket{u_m} = \sum_{h, \veck} \a{h, \, \veck} \rket{h, \veck; m},
\]
where the sum over $h$ is restricted to $ 0 \leq h \leq q-1.$ 
The condition $ P_{\ell-\half} \rket{u_m} = 0 $ has the same form as (\ref{P-0}).  
We study the coefficient $ \a{h,\vec0} $ with the largest possible value of $h.$ 
\begin{enumerate}
\renewcommand{\labelenumi}{(\roman{enumi})}
  \item $ m = 2p + 1 $ (odd) 
    \begin{itemize}
      \item $ p > q-1$ \\
        $ (h, k_0, \veck) = (q-1, 2(p-q)+3, \vec0) $ corresponds to the largest possible choice of $h. $ 
        The corresponding vector is $ \rket{q-1,\vec0; 2p} $ with the coefficents $ \a{q-1,\,\vec0} k_0 I_{\ell-\half} \mu. $ 
        Therefore $ \a{q-1,\, \vec0} = 0. $ 
      \item $ p \leq q-1 $ \\
        $ (h, k_0, \veck) = (p, 1, \vec0) $ corresponds to the largest possible choice of $h. $ 
        The corresponding term is $ \a{p,\,\vec0} k_0 I_{\ell-\half} \mu \rket{p,\vec0; 2p} $ so that 
        $ \a{p,\, \vec0} = 0. $ 
    \end{itemize}
    By the same argument as the proof of Theorem \ref{thm4} one may show that $ \a{h,\,\veck} = 0. $ 
    Therefore there are no singular vectors in $I^{\delta, \mu}$ at odd levels. 
 \item $ m = 2p $ (even)
   \begin{itemize}
     \item $ p > q-1 $ \\
       $ (h, k_0, \veck) = (q-1, 2(p-q+1), \vec0) $ corresponds to the largest possible choice of $h. $ 
       The corresponding vector $ \rket{q-1,\vec0; 2p-1} $ has nonvanishing coefficient so that 
       $ \a{q-1,\,\vec0} =0.$
     \item $ p \leq q-1 $ \\
       $ (h, k_0, \veck) = (p, 0, \vec0) $ corresponds to the largest possible choice of $h. $ 
       The corresponding vector has vanishing coefficient $ ( k_0 = 0 ) $ so that $ \a{p,\, \vec0} $ remains 
       undetermined. However, by a similar argument to the proof of Theorem \ref{thm4},
one requires 
       \beq
         2 \delta - 2(p-1) + \left( \ell + \half \right)^2 = 0, \label{delta-cond2}
       \eeq
       as the consistency condition. This condition is never satisfied since $ p \leq q-1 $ and 
       (\ref{delta-cond}) is supposed. This means that one may not determine $ \a{h,\,\veck}. $ 
   \end{itemize}
   Therefore there are no singular vectors in $I^{\delta, \mu}$ at even levels. 
\end{enumerate}
We thus arrive at the following theorem. 

\begin{thm} \label{thm5}
The irreducible highest weight modules of $ \cga $ for odd half-integer $ \ell $ with 
nonvanishing $ \mu $ are listed as follows:
 \begin{itemize}
   \item $\VM$ if $ \delta \neq q -1 - \half \left( \ell + \half \right)^2, $ 
   \item $ \VM / I^{\delta,\mu} $ if $ \delta = q -1 - \half \left( \ell + \half
\right)^2, $
 \end{itemize}
 where $ q \in \NN. $ All modules are infinite dimensional. 
\end{thm}

%%%%%%%%%%%%%%%%%%%%%%%%%%%%%%%%%%%%%%%%%%%%%%%%%%%%%%%%%%%%%%%%%%%%%%%%%%%%
%%%%%%%%%%%%%%%%%%%%%%%%%%%%%%%%%%%%%%%%%%%%%%%%%%%%%%%%%%%%%%%%%%%%%%%%%%%%
%%%%%%%%%%%%%%%%%%%%%%%%%%%%%%%%%%%%%%%%%%%%%%%%%%%%%%%%%%%%%%%%%%%%%%%%%%%%

\section{Kac determinant}
\label{Sec:KD}

Although we have already deduced a characterisation of the irreducible highest weight
modules of $\cga$ as presented in Theorem \ref{thm5}, we find that we are able to also
give the form of the Kac determinant corresponding to the subspace $\VM_m$ of the Verma
module at arbitrary level $m$.

We define a sesquilinear form $(,)$ on $\VM$ (Shapovalov form
\cite{Shap}) by setting
$$
\bracket{\delta,\mu}{\delta,\mu} \equiv (\hwv,\hwv) = 1,
$$
and
$$
(A\ket{x},B\ket{y}) = (\ket{x},\omega(A)B\ket{y}), \ \
\forall \ket{x},\ket{y}\in \VM,\ A,B\in \cga,
$$
where $\omega$ is an algebra anti-automorphism defined by
$$
\omega(P_j) = P_{2\ell-j},\ \omega(C)=H, \ \omega(H)=C,\ \omega(D)=D,\ \omega(M)=M.
$$
Note that $\omega$ is involutive, i.e. satisfies $\omega^2=$id. Generally, the form $(,)$
is hermitian. Restricting $(,)$ to the basis at level $m$ determined by the vectors
(\ref{mbasis}), we have the following.
\begin{lemma} \label{lem-sym}
The form $(,)$ is symmetric on the basis $\{ \ket{h,\veck ;m}\}$ of $\VM_m$.
\end{lemma}
\proof{
Since $\omega^2=$id, we clearly have $(\omega(A)x,y) = (x,\omega^2(A)y) = (x,Ay).$ Let
$A\hwv$ and $B\hwv$ be two vectors in the basis $\{\{ \ket{h,\veck ;m}\}$ of $\VM_m$. Then
\begin{eqnarray*}
(A\hwv,B\hwv)&=&(\hwv,\omega(A)B\hwv)\\ 
&=& \left\{ \begin{array}{rl} \alpha; & \mbox{ if }
\omega(A)B\hwv = \alpha \hwv\\ 0;& \mbox{ otherwise} \end{array}\right.
\\
&=&  \left(\omega(A)B\hwv,\hwv\right) = (B\hwv,A\hwv).
\end{eqnarray*}
Note that $\alpha\in\mathbb{R}$ in the above calculation. 
}

Given an ordered basis $\{ v_i\}$ of the level $m$ subspace $\VM_m$ of $\VM$, we define a
matrix whose entry in the $i$th row and $j$th column is the number $(v_i,v_j)$. Clearly
the null space of this matrix will lead to the set of vectors in $\VM_m$, called {\em null
vectors}, that are orthogonal to vectors in the basis. 
Elementary linear algebra then tells us that null vectors only exist if and only if the determinant of
the matrix is zero. This determinant is called the {\em Kac determinant at level $m$}.
In what follows, for $\ket{x}\in\VM$, we call $A\ket{x}$ a {\em descendent} of $\ket{x}$ for any
$A\in\cga^-$. The following lemma is included as a summary of some well-known results 
(e.g. see \cite{DMS}) concerning singular vectors and descendents of null vectors.

\begin{lemma} \label{kaclemma}
A singular vector is also a null vector, and descendents of a null vector are also
null vectors.
\end{lemma}
\proof{
Let $v_s$ be a singular vector at level $m$. Note that any vector $x$ at level $m$ can
be written as $x=Ax'$, for some $A\in\cga^-$ and $x'$ a vector whose level is less than
$m$. Note then that $\omega(A)\in\cga^+$, and
therefore $(x,v_s) = (x',\omega(A)v_s)=0.$ Hence $v_s$ is null. By a similar argument, if
$v$ is null, then $(x,Bv) = (\omega(B)x,v)=0,$ and hence $Bv$ is also null for all
$B\in\cga^-$.
}

The problem of determining singular vectors and finding highest weight submodules in the
Verma module is often associated with calculating the Kac determinant (e.g. see \cite{DMS}). Indeed,
the Kac determinant is usually used as a tool to undertake such analysis. Using
results of the previous sections, especially the completeness results of the singular
vectors (Theorem \ref{thm4}), we find that we can actually deduce a formula for the
Kac determinant. We include this in our paper more as a curious corollary to our main results,
rather than as a practical tool.

% In the discussion that follows, for convenience we consider bases of each level subspace
% $\VM_m$ comprising vectors of the form
% $$
% P_{\ell+\frac} C^h \prod_{j=1}^{\ell-\half} P_{\ell+\half + j}^{k_j} \hwv.
% $$
% This gives a slightly different basis of $\VM$ to the one introduced earlier, but we find
% it slightly more convenient to order the generators in terms of their {\em level weight}, i.e.
% how that generator affects the eigenvalue of $D$.

\subsection{Dimension of $\VM_m$}

Before looking at the details of the Kac determinant, we remark on the dimension of the
level subspace $\VM_m$. We note that the vectors in the basis $\{ \ket{h,\veck ;m}\}$ given in (\ref{mbasis}) 
for fixed level $m$ are
in one to one correspondence with the restricted set of integer partitions of the integer
$m$, with parts taken from the subset of integers 
\beq
\{2\} \cup \left\{2j+1\ | \ j=0,1,\ldots,\ell-\frac12\right\}.
\label{intparts}
\eeq
Observe that the vectors (\ref{mbasis}) in the basis of $\VM_m$ can be enumerated by the
nested labelling 
\begin{eqnarray}
&&\left\{ (h,k_1,\ldots,k_j,\ldots,k_{\ell-\frac12})\ \left|\ 0\leq h\leq \left\lfloor \frac{m}{2}\right\rfloor,\ 
0\leq k_{\ell-\frac12}\leq  \left\lfloor \frac{m-2h}{2(\ell-\frac12)+1}\right\rfloor,\ldots
\right.\right.\nn\\
&&
\qquad 
\ldots 0\leq k_{j}\leq  \left\lfloor\frac{m-2h-\sum_{n=j+1}^{\ell-\frac12}(2n+1)k_n}{2j+1}\right\rfloor,\
\ldots,\ \nn\\
&&
\qquad\qquad\ldots
\left.0\leq k_{1}\leq  \left\lfloor\frac{m-2h-\sum_{n=2}^{\ell-\frac12}(2n+1)k_n}{3}\right\rfloor\
\right\}.
\label{nestedlabels}
\end{eqnarray}
We therefore obtain the awkward yet explicit formula for the dimension of $\VM_m$, denoted $d^\ell_m$:
%\begin{eqnarray*}
$$
d^\ell_m =
\sum_{h=0}^{\left\lfloor \frac{m}{2}\right\rfloor}
\sum_{k_{\ell-\frac12}=0}^{\left\lfloor \frac{m-2h}{2(\ell-\frac12)+1}\right\rfloor}\ldots
\sum_{k_j
=0}^{\left\lfloor\frac{m-2h-\sum_{n=j+1}^{\ell-\frac12} (2n+1)k_n }{2j+1}\right\rfloor}
\ldots
%\\
%&&\qquad\ldots
\sum_{k_1=0}^{\left\lfloor\frac{m-2h-\sum_{n=2}^{\ell-\frac12} (2n+1)k_n }{3}\right\rfloor}
1.
$$
%\end{eqnarray*} 
For example, the first few cases of $\ell$ gives
\begin{eqnarray*}
d^{1/2}_m &=& \sum_{h=0}^{\left\lfloor \frac{m}{2}\right\rfloor}1,\\
d^{3/2}_m &=& \sum_{h=0}^{\left\lfloor \frac{m}{2}\right\rfloor}
\sum_{k_1=0}^{\left\lfloor \frac{m-2h}{3}\right\rfloor}1,\\
d^{5/2}_m &=& \sum_{h=0}^{\left\lfloor \frac{m}{2}\right\rfloor}
\sum_{k_2=0}^{\left\lfloor \frac{m-2h}{5}\right\rfloor}
\sum_{k_1=0}^{\left\lfloor \frac{m-2h-5k_2}{3}\right\rfloor}1,\\
d^{7/2}_m &=& \sum_{h=0}^{\left\lfloor \frac{m}{2}\right\rfloor}
\sum_{k_3=0}^{\left\lfloor \frac{m-2h}{7}\right\rfloor}
\sum_{k_2=0}^{\left\lfloor \frac{m-2h-7k_3}{5}\right\rfloor}
\sum_{k_1=0}^{\left\lfloor \frac{m-2h-7k_3-5k_2}{3}\right\rfloor}1.
\end{eqnarray*}

A more elegant approach is to characterise $d^\ell_m$ by the generating function \cite{Com,And}
\beq
  F^{\ell}(x) = \frac{1}{ 1-x^2 } \prod_{j=0}^{\ell-\half} \frac{1}{1-x^{2j+1}}, 
\label{d-genfun}
\eeq
in the sense that the coefficients of the formal power series are the numbers $d_m^\ell$,
i.e.
\[
 F^{\ell}(x)  
    = \sum_{m=0}^{\infty} d_m^{\ell} x^m.
\]
For example, using this generating function, we can apply standard combinatorial techniques \cite{Com}
to determine explicit formulas for the first few values of $\ell$:
\begin{eqnarray}
  & & d^{1/2}_m = \left\lfloor \frac{m+2}{2}  \right\rfloor,
  \label{dim-formula} 
\\[3pt]
  & & d^{3/2}_m = \left\lfloor \frac{m^2+6m+12}{12}  \right\rfloor,\nn
\\
  & & d^{5/2}_m = \left\lfloor \frac{2m^3+33m^2+162m+360}{360}  \right\rfloor, \nn \\[3pt]
  & & d^{7/2}_m = \left\lfloor \frac{m^4+36m^3+442m^2+2124m+5040}{5040}  \right\rfloor.
  \nn
\end{eqnarray}
Here we have used the notation $\lfloor x\rfloor = $max$\{N\in\Z\ | \ N\leq x\}$ is the usual floor function.

\subsection{Dependence on $\delta$ in the Kac determinant}

The result of Theorem \ref{thm1} establishes the existence of singular vectors for certain
values of $\delta$. Since all singular vectors are null vectors by Lemma \ref{kaclemma}, the 
Kac determinant must contain factors of the form $2\delta-2(q-1) + (\ell+\frac12)^2,$
arising from the existence condition for a singular vector at level $2q.$ 
For the Kac determinant at level $m$, such a factor must have algebraic 
multiplicity that is greater than or equal to the number of linearly independent
descendents of the singular vector at level $2q\leq m.$ We therefore have a lower bound on the 
algebraic multiplicity.

\begin{lemma} \label{lem-delord}
The Kac determinant at level $m$ contains the factor 
$$
\left(2\delta-2(q-1) + \left(\ell+\frac12\right)^2\right)^{d_{m-2q}^\ell},
$$
for every integer $q>0$ satisfying $m\geq 2q$.
\end{lemma}

Further to this, we can also see that only certain entries of the matrix of Shapovalov
forms involve factors of $\delta$, namely those entries whose Shapovalov form contains $C$ generators
in {\em both} basis vectors (e.g $(C^2P_{\ell+\frac12}^2\hwv,C^3\hwv)$ will be quadratic
in $\delta$). In the following, we use $a \sim \delta^h$ to indicate that $a$ is a
polynomial of degree $h$ in $\delta$.

\begin{lemma} \label{lem-delta}
Without loss of generality, let $h\leq h'$. Then $\bracket{h,\veck ; m}{h',\veck' ; m}$ is
either zero or
$$
\bracket{h,\veck ; m}{h',\veck' ; m} \sim \delta^h
$$
\end{lemma}
\proof{
It is a trivial matter to verify that some Shapovalov forms are zero. In the case it is
non-zero, we employ the commutations relations, particularly
$$
[H,C^{h'}] = -h' C^{h'-1}D + h'(h'-1)C^{h'-1},\ \ [D,P_n^k]=2k(\ell-n)P_n^k.
$$ 
We then have 
\begin{eqnarray*}
(C^h\mathbb{P}\hwv,C^{h'}\mathbb{P}'\hwv) &=&
(C^{h-1}\mathbb{P}\hwv,HC^{h'}\mathbb{P}'\hwv),\\
& = & -h'(C^{h-1}\mathbb{P}\hwv,C^{h'-1}D\mathbb{P}'\hwv)\\
&& \quad  +
h'(h'-1)(C^{h-1}\mathbb{P}\hwv,C^{h'-1}\mathbb{P}'\hwv),
\end{eqnarray*}
where $\mathbb{P}$ and $\mathbb{P}'$ represent some unimportant product of the $P_n$
generators. We arrive at the result by straightforward induction on $h$. Note also that we
lose no generality by setting $h\leq h'$ because $(,)$ is symmetric on the basis according to Lemma
\ref{lem-sym}. 
}

The significance of Lemma \ref{lem-delta} is that the diagonal entries of the matrix will
contain polynomials in $\delta$ of (non-strict) maximal degree in each row. Therefore, the
degree of the polynomial in $\delta$ occurring in the Kac determinant will be the sum of the 
degrees in the diagonal entries, which in turn is just the number of $C$ generators occurring in
all basis vectors of $\VM_m$.

In the following Lemma, we denote by $O^{2\ell}_n$ the number of integer partitions of $n$
comprising only odd parts no greater than $2\ell$. For convenience we adopt the convention
that $O^{2\ell}_0=1$.

\begin{lemma} \label{lem-sum}
$$
d_m^\ell = \sum_{n=0}^{\lfloor \frac{m}{2} \rfloor}O^{2\ell}_{m-2n}
$$
\end{lemma}
\proof{
It is well known (e.g. see \cite{Com,And}) that a generating function for the number of integer partitions of $n$
comprising only odd parts no greater than $2\ell$ is given by
$$
\prod_{j=0}^{\ell-\half} \frac{1}{1-x^{2j+1}} = \sum_{n=0}^\infty O^{2\ell}_nx^n,
$$
which occurs in the generating function (\ref{d-genfun}) for the dimensions of the level subspaces $\VM_m$.
Indeed, we see immediately that
\begin{eqnarray*}
 \sum_{m=0}^\infty d^\ell_m x^m = F^{\ell}(x) &=& \frac{1}{ 1-x^2 } \prod_{j=0}^{\ell-\half} \frac{1}{1-x^{2j+1}} \\
& = & \sum_{n=0}^\infty x^{2n} \sum_{t=0}^\infty O^{2\ell}_tx^t\\
& = & \sum_{n,t=0}^\infty O^{2\ell}_tx^{2n+t}\\
& = & \sum_{m=0}^\infty \sum_{n=0}^{\lfloor \frac{m}{2}\rfloor} O^{2\ell}_{m-2n}\ x^m
\mbox{ (setting $m=2n+t$)}
\end{eqnarray*}
from which the result follows.
}

It is clear that the number of vectors in the basis of $\VM_m$ containing $C^h$ must be
$O^{2\ell}_{m-2h}.$ It follows that the number of times the $C$ generators appear in
the basis vectors of $\VM_m$ is given by the expression
$$
\sum_{n=0}^{\lfloor \frac{m}{2} \rfloor} nO^{2\ell}_{m-2n}.
$$
As mentioned in the discussion following Lemma \ref{lem-delta}, this is precisely the
degree of the polynomial in $\delta$ that occurs in the Kac determinant. The result of
Lemma \ref{lem-sum} then implies that
$$
\sum_{j=0}^{\lfloor \frac{m}{2} \rfloor-1} d^\ell_{m-2(j+1)} = \sum_{n=0}^{\lfloor
\frac{m}{2} \rfloor} nO^{2\ell}_{m-2n},
$$ 
obtained by a simple rearrangement of the summation on the right hand side of the equation in Lemma \ref{lem-sum}. 
The left hand side coincides with the
sum of the powers of the factors obtained in Lemma \ref{lem-delord}. Therefore the result
of Lemma \ref{lem-delord} is not just a lower bound on the algebraic multiplicity, it is
precisely the algebraic multiplicity. We now have all
the pieces required to state the following.

\begin{thm} \label{thm-kacdel}
The Kac determinant at level $m$, denoted ${\cal D}^\ell_m$, is of the form  
$$
{\cal D}^\ell_m = f(\mu) \prod_{j=0}^{\lfloor \frac{m}{2}\rfloor-1}
\left( 2\delta-2j+\left(\ell+\frac12\right)^2 \right)^{ d^{\ell}_{m-2(j+1)} }.
$$
\end{thm}
Note that the function $f(\mu)$ remains as yet undetermined. At this stage we only
make the obvious point that the Kac determinant must depend on $\mu$. We now turn to
determining the form of $f(\mu)$.

\subsection{Dependence on $\mu$ in the Kac determinant}

It is convenient to introduce the notion of $\mu$-weight of a basis vector. Let the
basis $\{ \ket{h,\veck;m}\}$ of $\VM_m$ be denoted by $\gamma$. For any
$v\equiv\ket{h,\veck;m}\in\gamma$ we
define the {\em $\mu$-weight of $v$}, denoted $\rho_v$, as
\beq
\rho_v = m - 2\left(h+\sum_{j=1}^{\ell-\frac12} jk_j\right).
\label{muweight}
\eeq
Note that the $\mu$-weight of a vector is nothing more than the sum of powers of all the
$P_n$-type generators appearing in that basis vector. For example, in the case
$\ell=\frac52,$ consider the following three basis vectors at level 8: 
\begin{eqnarray*}
u &\equiv& \ket{1,\vec0;8}=CP_3^6\hwv,\\
v &\equiv&  \ket{1,\vece_1;8}=CP_3^3P_4\hwv,\\
w &\equiv&  \ket{0,\vece_1+\vece_2;8}=P_4P_5\hwv.\\
\end{eqnarray*}
Their $\mu$-weights are given by
$$
\rho_u=6,\ \ \rho_v = 4,\ \ \rho_w = 2.
$$
\begin{lemma} \label{lem-muweight}
For all $v,w\in\gamma,$ we have that either
$(v,w)=0$ or
$$
(v,w) =Z\mu^{\frac12\left( \rho_v + \rho_w \right)},
$$
for some $Z$ that has no dependence on $\mu$.
\end{lemma}
\proof{
As in the case of Lemma \ref{lem-delta}, it is easy to see that the Shapovalov forms
between some pairs of basis vectors are zero. In the case the form is non-zero, an
inductive proof by level can be used to prove the result. We outline such a calculation
here. Firstly the result is true at level 0 since $\bracket{\delta,\mu}{\delta,\mu}=1.$
We now look at the form 
$$
\bracket{h,\veck;m}{h',\veck';m},
$$
insisting only that $h\neq 0$, and assume the statement is true for all levels lower than this one.
Using the results of Lemma \ref{lemma2}, we have
\begin{eqnarray*}
\bracket{h,\veck;m}{h',\veck';m} &=& 
-\sum_{n=1}^{\ell-\frac12} k_n'\bracket{h-1,\veck;m-2}{h',k'+\vece_{n-1}-\vece_n;m-2}\\
&&
\quad+\frac12 k_0'(k_0'-1)\left( \left( \ell+\frac12 \right)! \right)^2\mu
\bracket{h-1,\veck;m-2}{h',\veck';m-2} \\
&&
\quad\quad+h'(m-h'-1-\delta)\bracket{h-1,\veck;m-2}{h'-1,\veck';m-2}.
\end{eqnarray*}
By the inductive assumption, we have
{\small
\begin{eqnarray*}
\bracket{h-1,\veck;m-2}{h',\veck'+\vece_{n-1}-\vece_n;m-2} & =&Z_1  
\mu^{\frac12\left(m-2-2\left(h-1+\sum_{j=1}^{\ell-\frac12}jk_j\right) +
m-2-2\left(h'+\sum_{j=1}^{\ell-\frac12}jk_j'+(n-1)-n\right) \right)}\\
&=& Z_1\mu^{\frac12\left(m-2\left(h+\sum_{j=1}^{\ell-\frac12}jk_j\right) +
m-2\left(h'+\sum_{j=1}^{\ell-\frac12}jk_j'\right) \right)},\\
\bracket{h-1,\veck;m-2}{h',\veck';m-2} & =&  
Z_2\mu^{\frac12\left(m-2-2\left(h-1+\sum_{j=1}^{\ell-\frac12}jk_j\right) +
m-2\left(h'+\sum_{j=1}^{\ell-\frac12}jk_j'\right) \right)}\\
&=& Z_2\mu^{-1+\frac12\left(m-2\left(h+\sum_{j=1}^{\ell-\frac12}jk_j\right) +
m-2\left(h'+\sum_{j=1}^{\ell-\frac12}jk_j'\right) \right)},\\
\bracket{h-1,\veck;m-2}{h'-1,\veck';m-2} & =& Z_3 
\mu^{\frac12\left(m-2-2\left(h-1+\sum_{j=1}^{\ell-\frac12}jk_j\right) +
m-2-2\left(h'-1+\sum_{j=1}^{\ell-\frac12}jk_j'\right) \right)}\\
&=& Z_3\mu^{\frac12\left(m-2\left(h+\sum_{j=1}^{\ell-\frac12}jk_j\right) +
m-2\left(h'+\sum_{j=1}^{\ell-\frac12}jk_j'\right) \right)},\\
\end{eqnarray*}
}
and so it is clear that in this case
$$
\bracket{h,\veck;m}{h',\veck';m}=Z 
\mu^{\frac12\left(m-2\left(h+\sum_{j=1}^{\ell-\frac12}jk_j\right) +
m-2\left(h'+\sum_{j=1}^{\ell-\frac12}jk_j'\right) \right)},
$$
as required, using the definition of $\mu$-weights. Note that in the above calculation,
$Z_1$, $Z_2,$ $Z_3$ and $Z$ are unimportant expressions with no $\mu$ dependence.
%Without loss of generality (i.e. by symmetry of $\bracket{\cdot}{\cdot}$), 
We now consider
$
\bracket{0,\veck_{(n)};m}{0,\veck';m},
$
where the notation $\veck_{(n)}$ implies that we impose the constraint
$k_0=k_1=\ldots=k_{n-1}=0$, and we apply this for each permissible $0\leq n\leq
\ell-\frac12$. In other words, we consider products of the form
$$
\left(\prod_{j=n}^{\ell-\frac12}P_{\ell+\frac12+j}^{k_j}\hwv ,
\prod_{j=0}^{\ell-\frac12}P_{\ell+\frac12+j}^{k_j'}\hwv \right),
$$
such that 
$$
\sum_{j=n}^{\ell-\frac12} (2j+1)k_j = m =  \sum_{j=0}^{\ell-\frac12}(2j+1)k_j'.
$$
Again, using the results of Lemma \ref{lemma2}, it is not difficult to verify that
$$
\bracket{0,\veck_{(n)};m}{0,\veck';m}
=k_n'I_{\ell-\frac12-n}\mu
\bracket{ 0,\veck_{(n)}-\vece_n;m-(2n+1) }{ 0,\veck'-\vece_n;m-(2n+1) }.
$$
By the inductive assumption, we have
$$
\bracket{ 0,\veck_{(n)}-\vece_n;m-(2n+1) }{ 0,\veck'-\vece_n;m-(2n+1) }\qquad\qquad\qquad\qquad
$$
$$
=Z' 
\mu^{\frac12\left( m-(2n+1)-2\left( \sum_{j=n}^{\ell-\frac12}jk_j-n\right) +
m-(2n+1)-2\left( \sum_{j=0}^{\ell-\frac12}jk_j' - n\right) \right)} 
$$
$$
 = Z'\mu^{-1+\frac12\left( m-2\left( \sum_{j=n}^{\ell-\frac12}jk_j \right)  +m-2\left(
\sum_{j=0}^{\ell-\frac12}jk_j \right) \right)}
$$
and hence
$$
\bracket{0,\veck_{(n)};m}{0,\veck';m} =Z''
\mu^{\frac12\left( m-2\left( \sum_{j=n}^{\ell-\frac12}jk_j \right)  +m-2\left(
\sum_{j=0}^{\ell-\frac12}jk_j \right) \right)}
$$
as required, with $Z'$ and $Z''$ being unimportant expressions with no $\mu$ dependence. 
The result is thus proved by induction, as we have now covered all
possibilities of the Shapovalov form using the fact that $\bracket{\cdot}{\cdot}$ is
symmetric on the basis, i.e. the result of Lemma \ref{lem-sym}.
}

Now that we have established how $\mu$ occurs in the Shapovalov forms, we are able to
deduce that $\mu$ occurs as a monomial in the Kac determinant. The simplest way of
viewing this is to recall the Leibniz formula for the determinant of a matrix.
For the purpose of readability, we set the size of the matrix to be $n$ ($=d^\ell_m$). 
Representing the basis vectors of $\VM_m$ as $\{ v_i\ |\ i=1,2,\ldots,n\}$, and  
recalling the Levi-Civita antisymmetric tensor 
$$
\epsilon^{i_1 i_2\ldots i_n} 
= \left\{ \begin{array}{rl} 
1;& (i_1, i_2,\ldots, i_n)\mbox{ even permutation of }(1,2,\ldots,n)\\ 
-1;& (i_1, i_2,\ldots, i_n)\mbox{ odd permutation of }(1,2,\ldots,n)\\ 
0;& \mbox{ otherwise,}
\end{array} \right.
$$
the Kac determinant can be expressed in the form
$$
{\cal D}^\ell_m  =  \sum_{i_1,i_2,\ldots,i_n=1}^n\epsilon^{i_1 i_2\ldots
i_n}(v_1,v_{i_1})(v_2,v_{i_2})\cdots(v_n,v_{i_n}).
$$
The result of Lemma \ref{lem-muweight} then implies that
$$
{\cal D}^\ell_m = \sum_{i_1,i_2,\ldots,i_n=1}^nZ_{i_1 i_2\ldots i_n}
\epsilon^{i_1 i_2\ldots i_n}
\mu^{\frac12\left(\rho_{v_1}+\rho_{v_{i_1}}\right)}
\mu^{\frac12\left(\rho_{v_2}+\rho_{v_{i_2}}\right)}
\cdots
\mu^{\frac12\left(\rho_{v_n}+\rho_{v_{i_n}}\right)},
$$
where the $Z_{i_1 i_2\ldots i_n}$ are expressions that have no dependence on $\mu$. Since
$\{ i_1,i_2,\ldots,i_n \}$ label all of the basis vectors, we must have
$$
{\cal D}^\ell_m = Z\mu^{\sum_{i=1}^n \rho_{v_i}}.
$$
The factor $Z$ has no dependence on $\mu$, but it may have dependence on $\delta$
depending on the level $m$. This dependence on $\delta$ in the Kac determinant has already
been  described in Theorem \ref{thm-kacdel}. 

Rather than give the final formula for the Kac determinant in terms of $\mu$-weights, which depends on knowledge of the basis, 
we seek a final expression in terms of $\ell$ and $m$ only. 
We denote by $e^\ell_m$ the sum of the $\mu$-weights over the basis, i.e.
\beq
e^\ell_m = \sum_{v\in\gamma}\rho_v,
\label{e-ell-m}
\eeq
where $\gamma$ represents the basis $\{ \ket{h,\veck;m}\}$ of $\VM_m$. In other words,
$e^\ell_m$ is the total number of $P_n$-type generators that occur in the basis $\gamma$,
which is furthermore equivalent to the total number of odd parts occurring in the
restricted integer partitions with parts taken from the subset of integers
(\ref{intparts}).
We can use the basis labelling given in the expression (\ref{nestedlabels}) along with the
definition of $\mu$-weight given in equation (\ref{muweight}) to rewrite equation
(\ref{e-ell-m}) as
$$
e^\ell_m =
\sum_{h=0}^{\left\lfloor \frac{m}{2}\right\rfloor}
\sum_{k_{\ell-\frac12}=0}^{\left\lfloor \frac{m-2h}{2(\ell-\frac12)+1}\right\rfloor}\ldots
\sum_{k_j
=0}^{\left\lfloor\frac{m-2h-\sum_{n=j+1}^{\ell-\frac12} (2n+1)k_n }{2j+1}\right\rfloor}
\ldots
%\\
%&&\qquad\ldots
\sum_{k_1=0}^{\left\lfloor\frac{m-2h-\sum_{n=2}^{\ell-\frac12} (2n+1)k_n }{3}\right\rfloor}
\left(  m-2\left( h+\sum_{j=1}^{\ell-\frac12}jk_j \right) \right).
$$
While this formula is explicit, it is rather crude. We can, however, adopt standard
techniques \cite{Com,And} to write down a generating function for the numbers $e^\ell_m$:
\beq
E^\ell(x) = \sum_{m=0}^\infty e^\ell_m x^m 
= \left( \sum_{i=0}^{\ell-\frac{1}{2}} \frac{x^{2i+1}}{1-x^{2i+1}}\right)
\frac{1}{1-x^2}\prod_{j=0}^{\ell-\frac{1}{2}}\frac{1}{1-x^{2j+1}}.
%= \left( \sum_{i=0}^{\ell-\frac{1}{2}}\frac{x^{2i+1}}{1-x^{2i+1}} \right)F^\ell(x).
\label{e-genfun}
\eeq

We have the following result for the form of the Kac determinant.

\begin{thm} \label{thm-kacdet}
The Kac determinant of $\cga$ at level $m$ is given by
$$
{\cal D}^\ell_m = C^\ell_m \mu^{e^\ell_m}\prod_{j=0}^{\lfloor \frac{m}{2}\rfloor-1}
\left( 2\delta-2j+\left(\ell+\frac12\right)^2 \right)^{ d^{\ell}_{m-2(j+1)} },
$$
for some constant $C^\ell_m,$ and where $d^\ell_{m-2(j+1)}$ and $e^\ell_m$ are determined
by the generating functions (\ref{d-genfun}) and (\ref{e-genfun}) respectively.
\end{thm}

Note that in the case $\ell=1/2$, we have
\begin{eqnarray*}
e^{1/2}_m &=& \sum_{h=0}^{\left\lfloor \frac{m}{2}\right\rfloor}(m-2h) 
= \left( m-\left\lfloor \frac{m}{2} \right\rfloor  \right)
  \left( \left\lfloor \frac{m}{2} \right\rfloor +1 \right)\\
&=&\left\{ 
\begin{array}{rl} 
\frac14m(m+2);& m\mbox{ even}\\
\frac14(m+1)^2;& m\mbox{ odd}.
\end{array} \right.
\end{eqnarray*}
and from equation (\ref{dim-formula}), 
$$
d^{1/2}_{m-2(j+1)} = \left\lfloor \frac{m-2(j+1)+2}{2} \right\rfloor = \left\lfloor
\frac{m-2j}{2} \right\rfloor=
\left\lfloor \frac{m}{2} \right\rfloor-j.
$$
The result of Theorem \ref{thm-kacdet} therefore confirms the conjectured form of the Kac determinant for
$\ell=\frac12$ that was presented in \cite{DoDoMr}.

\section{Concluding remarks}
%\label{Sec:CR}

The main results of this paper are presented in Theorem \ref{thm5} and Theorem
\ref{thm-kacdet}. Theorem \ref{thm5} is the culmination of our study of singular vectors
in the Verma module of $\cga$, and classifies all the irreducible highest weight modules.
Theorem \ref{thm-kacdet} gives the form of the Kac Determinant, that ultimately comes
about as a result of the study of singular vectors. For the special case $\ell=1/2$, we
have demonstrated explicitly that the conjectured form of the Kac determinant presented in
\cite{DoDoMr} is a special case of the result of Theorem \ref{thm-kacdet}.

%%%%%%%%%%%%%%%%%%%%%%%%%%%%%%%%%%%%%%%%%%%%%%%%%%%%%%%%%%%%%%%%%%%%%%%%%%%%%%
%
%  Acknowledgements
%
%%%%%%%%%%%%%%%%%%%%%%%%%%%%%%%%%%%%%%%%%%%%%%%%%%%%%%%%%%%%%%%%%%%%%%%%%%%%%%
%
\section*{Acknowledgements}

N.A. is supported by a grants-in-aid from JSPS (Contract No.23540154). Most of this work
was done during a visit of P.S.I to Osaka Prefecture University (OPU).
P.S.I. would like to thank OPU for their hospitality during his stay.

%%%%%%%%%%%%%%%%%%%%%%%%%%%%%%%%%%%%%%%%%%%%%%%%%%%%%%%%%%%%%%%%%%%%%%%%%%%%%%
%
%  References
%
%%%%%%%%%%%%%%%%%%%%%%%%%%%%%%%%%%%%%%%%%%%%%%%%%%%%%%%%%%%%%%%%%%%%%%%%%%%%%%
%

\end{document}